# Hydrodynamic model of the collective electron resonances in $C_{60}$ fullerene

V.B. Gildenburg [1,2,*], I.A. Pavlichenko [1,2]

[1] University of Nizhny Novgorod, Nizhny Novgorod 603950, Russia
[2] Institute of Applied Physics, Russian Academy of Sciences, Nizhny Novgorod 603950, Russia
[*] gil@appl.sci-nnov.ru

The polarization-response spectrum of the fullerene $C_{60}$ modeled as a homogeneous spherical plasma shell is calculated in the framework of the hydrodynamic approach allowing for the spatial dispersion caused by the Fermi-distributed valence electrons. The dipole eigenoscillations spectrum of the shell is found to contain a series of plasmons distinguished by the frequency and the radial structure. The first two of them (whose structures for $C_{60}$ is the subject of discussion up to now) pass to the lower and higher *surface* plasmons of the plasma shell if its thickness is much larger than Tomas-Fermi length. However, under parameters values corresponding to the $C_{60}$ molecule, when these lengths are of the same order, both these plasmons (providing the main contribution to the fullerene absorption spectrum) are found to be actually *volume* ones in their spatial structure, and the frequency of the higher of them becomes larger than the plasma frequency (as with all the higher volume plasmons). The resonance curve of the fullerene absorption cross-section calculated on the basis of the developed model with allowance for the *surface losses* caused by the reflection of electrons at the shell boundaries agrees satisfactorily with the experimental data.

Studies on the dynamical polarizability of the spherical plasma shells interacting with the fields of different external sources are of interest in connection with the problems of the observed scattering and absorption spectra identification of the fullerenes (and, in particular, $C_{60}$ molecule), displaying oscillatory features of plasma-like systems due to a great number of the valence (delocalized) electrons [1-18]. Related circle of problems arises also when studying metamaterials made on the basis of shell-like structures (for example, semiconductor quantum dots or metal-dielectric nanoparticles) [19-27]. The studies described in the references above have allowed to reveal a number of important features and characteristics of the fullerene polarization processes. There was predicted [1] and confirmed experimentally [2-5] the existence of the so-called giant resonance of $C_{60}$ molecule in the frequency range ~ 20 – 40 eV, associated with the excitation of the dipole surface plasmons. The plasmon resonances spectra of this molecule were studied in different aspects: based on the quantum [5-11] and semi-classical [12-18] approaches (including the nonlinear effects in the processes of the one-photon and multi-photon absorption [9]), with the dipole and high multipoles excitations taken into account [6-8,14-18], under different manifestation conditions (excitation by the optical radiation [5, 9-18] and moving electrons [6-8]), and different approximations of electron density distribution (homogeneous [5-10,13,14] or inhomogeneous shells [11,12] and 2D spherical films [15-18]). Nevertheless, the carried out researches have not yet brought the complete enough and striking model agreeing well with experiment. In particular, so far it is not clear to the end the full frequency spectrum, the spatial structure of the main plasmons, and mechanisms determining their damping constants. One of the difficulties is associated with in the description of the plasma-field interaction under conditions of the strong polarization nonlocality (termed "strong spatial dispersion" in the plasma electrodynamics). The nonlocal effects are responsible for the spectral shifts of collective electron excitations (surface and volume plasmons) and can affect drastically their spatial structure; in fact, they determine also the dissipations mechanisms, and the corresponding damping constants, which were chosen empirically in the most fulfilled studies.

In this paper, the polarization response spectrum of the plasma shell structure, that models the fullerene $C_{60}$, is calculated based on the hydrodynamic approach, supplemented by some results of kinetics description for evaluation of the energy loss parameters. This approach seems to permit an adequate account of the spatial dispersion caused by the Fermi motion of the valence electrons and proper analysis of its associated spectrum features.

We consider the homogeneous plasma shell as a simple model of $C_{60}$ molecule: it contains 240 valence electrons that are assumed to be free ones and distributed uniformly between two concentric spheres of radii $a = 2.5$ Å and $b = 4.6$ Å (see Refs. [5,11-13]) with the density $N = 240[(4\pi/3)(b^3 - a^3]^{-1} = 6.9 \times 10^{23}$ cm$^{-3}$. The shell is placed in the alternating external electric field $\mathbf{E}_0 = \mathbf{e}_z E_0 \exp(-i\omega t)$. In the frequency range of interest, the wavelength is much greater than the shell size ($kb \ll 1$, $k = \omega/c$, $c$ is the light velocity in the vacuum) to allow for the potential description of the field inside and around the shell ($\mathbf{E} = -\nabla\varphi \cdot \exp(-i\omega t)$) reducing the calculation of the field to the boundary problem for the potential $\varphi(\mathbf{r})\exp(-i\omega t)$. In the hydrodynamic approximation, the material equation relating complex amplitudes of the displacement ($\mathbf{D}\exp(-i\omega t)$) and the electric field ($\mathbf{E}\exp(-i\omega t)$) in homogeneous degenerate plasma can be written as [28-31]

$$\mathbf{D} = \varepsilon\mathbf{E} + \frac{3}{5}\frac{V_F^2}{\omega^2 + i\omega\nu}\nabla(\nabla \cdot \mathbf{E}). \tag{1}$$

Here, $V_F = (3\pi^2 N)^{1/3}\hbar/m$ is the Fermi velocity, $\varepsilon = 1 - \omega_p^2/(\omega^2 + i\omega\nu)$ is the dielectric function of the free electron gas, given by Drude's theory, $\omega_p = (4\pi e^2 N/m)^{1/2}$ is the plasma frequency, $e$ and $m$ are the electron charge and mass respectively, $\hbar$ is the Planck constant, $\nu$ is the dissipation parameter ("effective electron collision frequency") allowing, in principle, for the different loss mechanisms). With Eq. (1) taken into account, Maxwell equation $\nabla \cdot \mathbf{D} = 0$ allows for representation of the electric potential in plasma as a sum of two terms, that satisfy independent equations [29,31-33]:

$$\varphi = \varphi_t + \varphi_p, \quad \Delta\varphi_t = 0, \quad \Delta\varphi_p + k_p^2\varphi_p = 0, \tag{2}$$

where $k_p = \sqrt{5\varepsilon/3}\,\omega/V_F$ is the wave number of the longitudinal wave. Outside the plasma, the potential satisfies the Laplace equation

$$\Delta\varphi = 0. \tag{3}$$

In the spherical coordinates $r, \vartheta$ (radius $r$ is distance from the center of the shell, $\vartheta$ is the angle between the radius and $z$-axes), the solution of the Eqs. (2) and (3) that is finite at the origin of coordinate an at infinity, has the form $\varphi(r,\vartheta) = \Phi(r)\cos\vartheta$, where the function $\Phi(r)$ for the different regions is given by

$$r < a: \quad \Phi(r) = C_0 r, \tag{4}$$

$$a < r < b: \quad \Phi(r) = C_1 r + C_2 r^{-2} + C_3 j_1(k_p r) + C_4 y_1(k_p r), \tag{5}$$

$$r > b: \quad \Phi(r) = -E_0 r + P r^{-2}. \tag{6}$$

Here, $j_1(\xi)$ and $y_1(\xi)$ are the first-order spherical Bessel function of the first and second kind, respectively, the constants $C_0$, $C_1$, $C_2$, $C_3$, $C_4$, and $P$ (the last of them is the dipole moment of the shell) can be found from the boundary condition at $r=a$ and $r=b$, that are the continuity of the potential and the normal components of the displacement and the electric field (the latter condition implies the absence of electron current density at the plasma boundaries; see also Refs. [29, 31-33]). Because the expressions for these constants are too cumbersome, we present here only most important of them, namely, the dipole moment:

$$P = \left\{ \frac{3\varepsilon}{\Delta}\left[ 2\varepsilon+1-2(\varepsilon-1)\frac{f_1+q_a^3}{g}+2\eta^3(1-\varepsilon)\left(1-\frac{f_2+q_b^3}{g}\right)\right] - 1\right\} \frac{b^3 E_0}{2}, \quad (7)$$

where

$$\Delta = \left[\varepsilon+2+2(\varepsilon-1)\frac{f_1-q_a^3}{g}\right]\cdot\left[2\varepsilon+1-2(\varepsilon-1)\frac{f_2+q_a^3}{g}\right]$$
$$-2\left(1-\frac{f_2+q_b^3}{g}\right)\cdot\left(1+2\frac{f_1-q_b^3}{g}\right)\cdot(1-\varepsilon)^2\eta^3, \quad (8)$$

$$f_{1,2} = (2q_a q_b - q_{a,b}^2 + 2)\sin q \pm [q_{b,a}(q_{a,b}^2-2)+2q_{a,b}]\cos q,$$
$$g = 2[q_a(q_b^2-2)-q_b(q_a^2-2)]\cos q - [4q_a q_b + (q_a^2-2)(q_b^2-2)]\sin q, \quad (9)$$
$$q_a = k_p a, \ q_b = k_p b, \ q = k_p l, \ \eta = a/b, \ l = b-a.$$

The found solution generalizes the one obtained for the plasma shell without allowance for the spatial dispersion [12,13] and passes to it in the formal limit $V_F \to 0$, $k_p \to \infty$.

Putting $\Delta = 0$ in Eq. (7), we find the characteristic equation determining, along with the above formulas for $\varepsilon(\omega)$ and $k_p(\omega)$, the spectrum of complex eigenfrequencies of shell dipole plasmons. In the absence of the spatial dispersion this spectrum contains two surface plasmons; at small losses, the real parts of their eigenfrequencies and the corresponding resonance values of the dipole moments of the shell are as follows:

$$\omega_{0,1}^2 = \omega_p^2\left(\frac{1}{2}\mp\frac{1}{6}\sqrt{1+8\eta^3}\right), \quad (10)$$

$$P_{0,1} = \frac{ib^3 E_0 \omega_{0,1}(\omega_p^2-\omega_{0,1}^2)(2\omega_p^2-3\omega_{0,1}^2)}{2\nu(\omega_p^2-2\omega_{0,1}^2)}. \quad (11)$$

The changes in the shell plasmon spectrum, caused by the spatial dispersion effects are in principle the same as for the solid spherical cluster [29, 31-33] and they are determined by the relation between the Thomas-Fermi length $\lambda_{TF} = V_F/\omega_p$ and the characteristic structure dimensions. There are two main such effects: (i) a frequency blueshift of surface plasmons – that is so called "dimension effect" (that is noticeably pure classical here, but not quantum one); (ii) enrichment of the resonant spectrum by the series of the volume plasmons lying in the region of $\omega > \omega_p$. We note, that the analogues approach to the problem considered was developed in Ref. [14], where the eigenfrequencies of the lowest (with $\omega < \omega_p$ only) multipole modes were calculated, and the questions below being the main subject of this work (the spatial structures of different modes, their damping constants, and excitation by the external field, as well as the volume plasmons existence) were not addressed.

Results of numerical calculations of the resonance curve $P(\omega)$ based on the obtained general solution as applied to the above parameters of the fullerene $C_{60}$ are presented below, but first, to elaborate some qualitative understanding the nature of different resonances, it seems reasonable to trace changing the full spectrum and space structure of collective excitation with spatial dispersion parameter $u = \lambda_{TF}/l = V_F/(\omega_p l)$. Dependences of the plasmon eigenfrequencies and the space charge distribution for the corresponding plasmons in the shell on this parameter are presented at the above shell sizes with no losses (at $\nu = 0$) in Figs.1-3 (the thin vertical line on Fig.1 corresponds to the value $u = 0.314$ for the fullerene).

In the absence of spatial dispersion (at $V_F = 0$) the frequencies of the surface plasmons and corresponding resonant values of the dipole moments at small losses, as derived from Eqs.(7) – (10), are

$$\frac{\omega_0}{\omega_p} \approx 0.5, \quad \frac{\omega_1}{\omega_p} \approx 0.87, \quad P_0 = \frac{0.45 i E_0 b^3 \omega_p}{\nu}, \quad P_1 = \frac{0.06 i E_0 b^3 \omega_p}{\nu}, \quad (12)$$

that is the higher-frequency resonance is (at the same losses) approximately one order weaker than the lower-frequency one. All the formally calculated frequencies of the volume plasmons are the same at $V_F = 0$ and coincident, as seen in Fig. 1, with the plasma frequency $\omega_p$, but the resonance in this case is absent: the dipole moment $P_n = -b^3 E_0/2$. This, in particular, implies, that the description of the volume plasmons excitation is possible only with spatial dispersion taken into account. With growing $V_F$, all the eigenfrequencies increase: the frequency of the lower mode $\omega_0$ grows comparatively slowly and remains smaller than $\omega_p$; the frequency of the higher surface plasmon $\omega_1$ passes through $\omega_p$ at some $V_F$ dependent on the shell dimensions and continues to grow together with the eigenfrequencies of other volume plasmons $\omega_n$ ($n = 2,3,4,...$) that are defined approximately by the condition $k_p l \approx \pi n$, whence it follows

$$\omega_n^2 \approx \omega_p^2 + \frac{3\pi^2 n^2 V_F^2}{5l^2}. \quad (13)$$

(this formula become exact in the limit $l/a \to 0$).

The structure of difference plasmon modes can be inferred from Figs. 2 showing space charge distributions in several first plasmons calculated numerically in the absence of dissipation at the fullerene parameters. Changing of these distributions with the Fermi's velocity is illustrated by the curves on Fig. 3.

It should be stressed that the separations of the plasmons into the "surface" and "volume" ones is justified only under conditions $\lambda_{TF} \ll a, b-a$, when the space distribution of their oscillating space charge density $\rho_n = k_p^2 \varphi_{pn}/(4\pi)$ corresponds to their names. In the surface plasmons, the space charge appears to be almost surface one and it is localized near the plasma-vacuum interfaces (Fig. 3 a, b, dashed curves), meanwhile the volume plasmons represent, in fact, the standing longitudinal waves (or the waves of a space charge with different number of the half-waves on the shell thickness) whose electron density variations occupy the whole plasma volume (Fig. 3 c). In the case of a strong spatial dispersion, when the above inequalities are not satisfied, such separation loses its meaning because the space charge distribution is of a volume character for both the types of plasmons (Fig. 2 a, b, c and Fig. 3, solid curves). In this context, a discussion of the surface/volume character of the fullerene plasmons [5,11,34,35] seems to be rather pointless. Both the first and second branches ($n = 0$ and $n = 1$) trace back to the surface plasmons in the media with local polarizability (where the nonlocality parameter $u \ll 1$) but under fullerene conditions (where $u \sim 1$) the corresponding plasmons transform actually to the volume ones (see Figs. 1-3). Moreover, the second of them attains the eigenfrequency lying in the region $\omega > \omega_p$, as with "genuine" volume plasmons.

Actually, with increasing the spatial dispersion parameter $u$ all the plasmons convert to the volume ones, but they do not change the character of their symmetry, determined by the mode number $n$ that is equal to the number of zeros of the space charge function $\rho_n(r)$ (at the given $\vartheta$) on the interval $a \leq r \leq b$. For the even $n$, the signs of this function are the same at $r = a$ and $r = b$, for the odd $n$ they are opposite. At $n \geq 2$ these function are approximately proportional to $\cos[n\pi(r-a)/l]$. At the thin enough shell ($l = b - a < a$), the important distinctive marker for the first two plasmons is also the difference in primary directions of the polarization currents: in the lowest mode ($\omega = \omega_0$) the meridian-oriented fields and currents (with components $E_\vartheta$, $j_\vartheta$) prevail, while in the higher one ($\omega = \omega_1$), as well as in the volume plasmons, the radial components ($E_r$, $j_r$) dominate [13].

The dipole resonance peaks of different plasmons and the relation between them in the common spectrum of polarizability $P(\omega)$ depend strongly on the dissipation parameter $\nu_n$ and the resonance excitation factor that can be determined as the coefficient $\beta_n$ in the expression for the imaginary part of the dipole moment $P_n$ at the $n$-s plasmon resonance:

$$\mathrm{Im}\, P_n = b^3 E_0 \frac{\beta_n \omega_p}{\nu_n}, \qquad (14)$$

and characterizes, in fact, the coupling between the plasmon and the external field. In the absence of the spatial dispersion for the lower and higher plasmons, this factor, as it follows from Eq. (12), is equal to 0.45 and 0.06, respectively. Its dependence on the spatial dispersion parameter for different plasmons is shown in Fig. 4. As seen, the values $\beta_n$ for the even volume plasmons (at $n \geq 2$) are much smaller then those for the odd ones and decrease fast with the number of plasmon mode.

Among different possible dissipation mechanisms, the main role in the studied system is played evidently by the collisionless one, (i.e., the Landau damping) and we can take it into account in the framework of the hydrodynamic approach by estimating the parameter $\nu$ on the basis of some simplified kinetic models. For the collective electron excitations in the bounded nanoplasma with characteristic dimension $L$ that determines the mean free path of electrons in the direction of its oscillations (for example, it is the radius of the solid spherical cluster or the thickness of the plane layer), the parameter of collisionless dissipation $\nu$ evaluated in the framework of such models commonly appears to be approximately equal to the inverse characteristic free-flight time of electrons: $\nu = A V_F / L$, where $A$ is the coefficient of the order of unity [29,31,36,37]. As applied to the shell resonances, it is naturally to set $L = a$ for the lowest plasmon with $n = 0$ (where the mean free pass length of the electrons in the direction of its meridional oscillations is of the order of the smaller radius) and $L = l$ for the higher plasmon (transformed to the volume one with growing $V_F$) and several low-lying volume modes (where this length in the direction of radial oscillations is the shell thickness). As a result, the dissipation parameter turns out to be comparatively large (at $A = 1$, we have $\hbar\nu = 9$ eV and 15 eV for the first and second plasmons, respectively), so that, in view of fast decreasing the excitation factors $\beta_n$ with mode number, only these two modes are found to be appreciably presented in the polarizability spectrum of the fullerene. It should be noted, however, that this situation, can be changed if the collective electron oscillations are excited by other means, for example, by moving electrons or ions, so that in the EELS spectra of fullerene and its compounds, the resonances of highest volume plasmons, can play, possibly, a perceptible role.

The resonance curves in Fig. 5 shows the frequency dependences of the absorption cross-section of the shell

$$\sigma_a = \frac{4\pi\omega \operatorname{Im} P}{cE_0}, \qquad (15)$$

which were calculated using Eqs. (7)-(9) at the above parameters for the fullerene $C_{60}$ (solid curve) and obtained in the experiments described in Ref [3] (dashed and dotted curves). The main peak on the curves correspond to the resonance of the lowest plasmon (marked by the index $n = 0$ in our notations). The slight maxima at the flattened right slope of the main peak is caused by the next plasmon (with index $n = 1$). Taking into account a comparatively simplicity of the model used, the calculated resonance curve may be considered satisfactorily consistent with the experimental ones. Analogues theoretical results were obtained also based on more complicated quantum methods (TDLDA in Ref. [5] and TDTF in Ref. [14]). However, in these works the interpretation of results concerning the nature of plasmons was not based on the direct calculation of their structure, and it differs considerably from our work.

To conclude, we have studied the polarizability of $C_{60}$ fullerene in the frequency range corresponding to its collective (plasmon) resonances. In the framework of the comparatively simple hydrodynamic approach, we have found the plasmon spectrum of a spherical plasma shell modeling this molecule with allowance for the spatial dispersion caused by the Fermi motion of valence electrons. This spectrum is characterized by the series of collective dipole modes distinguished by the eigenfrequencies and the radial distributions of the vibrating electron density. The oscillating potential and space charge in each mode are defined by a superposition of the solutions of the Laplace equation and the Helmholtz equation for the plasma waves. The first two plasmons convert to the lower and higher surface plasmon of the shell when decreasing Fermi's velocity, but under parameter values corresponding to the sizes of fullerene $C_{60}$ their vibrating space charge is distributed over the whole volume of the shell. The damping constants (line width's) of plasmons are determined in our model by the so-called surface losses, and they are equal approximately to the characteristic frequency of the electron collision with the shell boundaries. This losses lead to strong damping of all highest modes (volume plasmons). As a result, only first two plasmon can contribute really to the resonance spectrum of the polarization response of $C_{60}$ interacting with the external optical field, so that the resonance curve of the absorption cross-section we have calculated on the basis of our model contains, in accordance with the experimental data, the one large peak corresponding to the excitation of the lowest plasmon (~ 20 eV) and slight peak (more likely manifested as flattened plateau at the right slope of the main peak) associated with the second plasmon (~40 eV).

The development of the analytical hydrodynamic model and derivation of analytical formulas were supported by the Russian Science Foundation (Grant No. 17-12-01574). Elaboration of a numerical code and numerical calculations of the fullerene plasmon spectra were performed with the support of the Russian Government (Agreement No. 14.B25.31.0008).

1. G. F. Bertsch, A. Bulgac, D. Toma´nek, and Y.Wang, Phys. Rev. Lett. **67**, 2690 (1991).
2. I.V. Hertel, H. Steger, J. de Vries, B. Weisser, C. Menzel, B. Kamke, and W. Kamke, Phys. Rev. Lett. **68**, 784 (1992).
3. H. Yagi, K. Nakajima, K.R. Koswattage, K. Nakagawa, C. Huang, Md.S.I. Prodhan, B.P. Kafle, H. Katayanagi, K. Mitsuke, H. Yagi, K. Nakajima, K.R. Koswattage, Carbon **47**, 1152 (2009).
4. C.M. Thomas, K.K. Baral, N.B. Aryal, M. Habibi, D.A. Esteves-Macaluso, A.L.D. Kilcoyne, A. Aguilar, A.S. Schlachter, S. Schippers, A. Müller, and R.A. Phaneuf, Phys. Rev. A **95**, 053412 (2017).
5. S.W.J. Scully, E.D. Emmons, M.F. Gharaibeh, and R.A. Phaneuf, Phys. Rev. Lett. **94**, 065503 (2005).

6. L.G. Gerchikov, A.V. Solov'yov, J.P. Connerade, W. Greiner, J. Phys. B: At. Mol. Opt. Phys. **30**, 4133 (1997).
7. A.V. Verkhovtsev, A.V. Korol, A.V. Solov'yov, P. Bolognesi, A. Ruocco and L. Avaldi, J. Phys. B: At. Mol. Opt. Phys. **45**, 141002 (2012).
8. C.Z. Li, Z.L.Miskovic, F.O.Goodman, Y.N.Wang, J. Appl. Phys. **113**,184301 (2013).
9. J.P. Connerade and A.V. Solov'yov, Phys. Rev. A **66**, 013207 (2002).
10. J. Choi, E. Chang, D.M. Anstine, M. El-Amine Madjet, and H. S. Chakraborty, Phys. Rev. A **95** , 023404 (2017).
11. D.I. Palade, and V. Baran, J. Phys. B: At. Mol. Opt. Phys. **48** 185102 (2015).
12. D. Ostling, S.P. Apell, G. Mukhopadhyay, A. Rosen, J. Phys. B: At. Mol. Opt. Phys. **29**, 5115 (1996).
13. Ph. Lambin, A. A. Lucas, and J.-P. Vigneron, Phys. Rev. B, **46**, 1794 (1993).
14. M. Michalewicz, M.P.Das, Solid State Commun. **84**, 1121 (1992).
15. A. Moradi, Optik **123**, 325 (2012).
16. A. Moradi, Solid State Commun. **192**, 24 (2014).
17. A. Moradi, Phys. Plas. **23**, 062120 (2016).
18. A. Moradi, Optik **143**, 1 (2017).
19. E. Prodan, C. Radloff, N.J. Halas, P. Nordlander, Science **302**, 419 (2003).
20. P.K. Jain and M.A. El-Sayed, The Journal of Physical Chemistry C **111**, 17451 (2007).
21. R. Bardhan, N.K. Grady, T. Ali, and N.J. Halas, ACS Nano **4**, 6169 (2010).
22. J.M. Luther, P.K. Jain, T. Ewers, and P. Alivisatos, Nature Matter **10**, 361 (2011).
23. W. Liu, A.E. Miroshnichenko, D.N. Neshev, and Y.S. Kivshar, ACS Nano **6**, 5489 (2012).
24. J. Zhu, J.-J. Li, L. Yuan, and J.-W. Zhao, The Journal of Physical Chemistry C **116**, 11734 (2012).
25. F. Shirzaditabar, M. Saliminasab, and B.A. Nia, Phys. Plas. **21**, 072102 (2014).
26. A. Moradi, Phys. Plas. **22**, 042105 (2015).
27. A.V. Zasedatelev, T.V. Dubinina, D.M. Krichevsky, V.I. Krasovskii, V.Y. Gak, V.E. Pushkarev, L.G. Tomilova, and A.A. Chistyakov, The Journal of Physical Chemistry C **120**, 1816 (2016).
28. V.L. Ginzburg, The propagation of electromagnetic waves in plasmas. International Series of Monographs in Electromagnetic Waves, Oxford: Pergamon (1970).
29. V.B. Gildenburg, V.A. Kostin, I.A. Pavlichenko, Phys. Plas. **18**, 092101 (2011).
30. N.A. Mortensen, Photonics and Nanostructures - Fundamentals and Applications **11**, 303 (2013).
31. .B. Gildenburg, V.A. Kostin, I.A. Pavlichenko, Phys. Plas. **23**, 032120 (2016).
32. V.B. Gildenburg and I.G. Kondrat'ev, Radiotekhnika i Elektronika **10**, 658 (1965) [in Russian]; translation: Radio Eng. Electr. Phys. – USSR **10**, 560 (1965).
33. R. Ruppin, Phys. Rev. Lett. **31**, 1434 (1973).
34. A.V. Korol and A.V. Solov'yov, Phys. Rev. Lett. **98**, 179601 (2007).
35. S.W.J. Scully, E.D. Emmons, M.F. Gharaibeh, R.A. Phaneuf, A.L.D. Kilcoyne, A.S. Schlachter, S. Schippers, A. Mü̈ller, H.S. Chakraborty, M.E. Madjet, and J.M. Rost, Phys. Rev. Lett. **98**, 179602 (2007).
36. H. Hövel, S. Fritz, A. Higler, U. Kreibig, and M. Vollmer, Phys. Rev. B **48**, 18178 (1993).
37. C. Yannouleas and R.A. Broglia, Ann. Phys. **217**, 105 (1992).

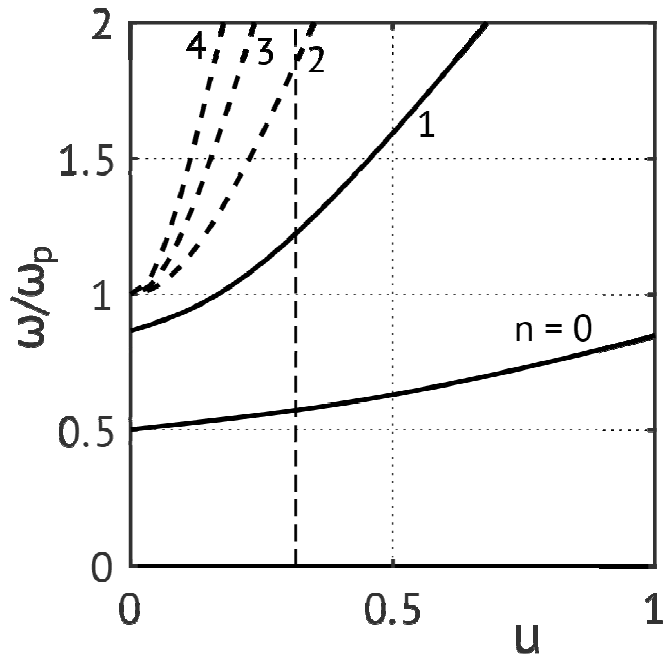

Fig. 1. Plasmon eigenfreqencies $\omega_n$ of fullerene $C_{60}$ with no losses as a function of the nonlocality parameter $u = V_F /(\omega_p l)$ for the first five modes ($n = 0, 1, 2, 3, 4$; $V_F$ is Fermi's velocity, $\omega_p$ is the plasma frequency, $l = b - a$ is the shell thickness). The vertical dashed line $u = 0.314$ corresponds to the parameters of $C_{60}$.

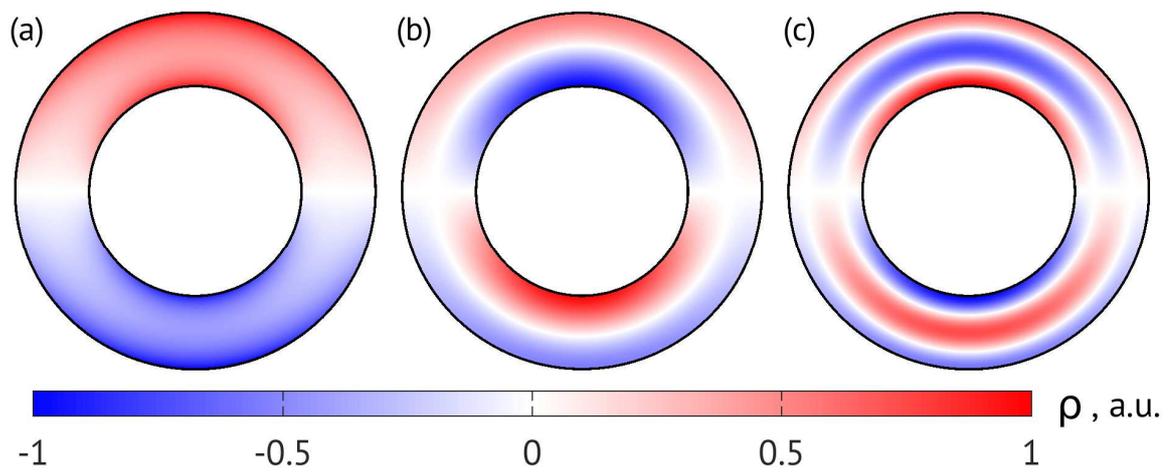

Fig. 2. Space charge density distributions $\rho_n(\mathbf{r})$ for the first three plasmon eigenmodes in $C_{60}$: (a) $n=0$, (b) $n=1$, (c) $n=2$.

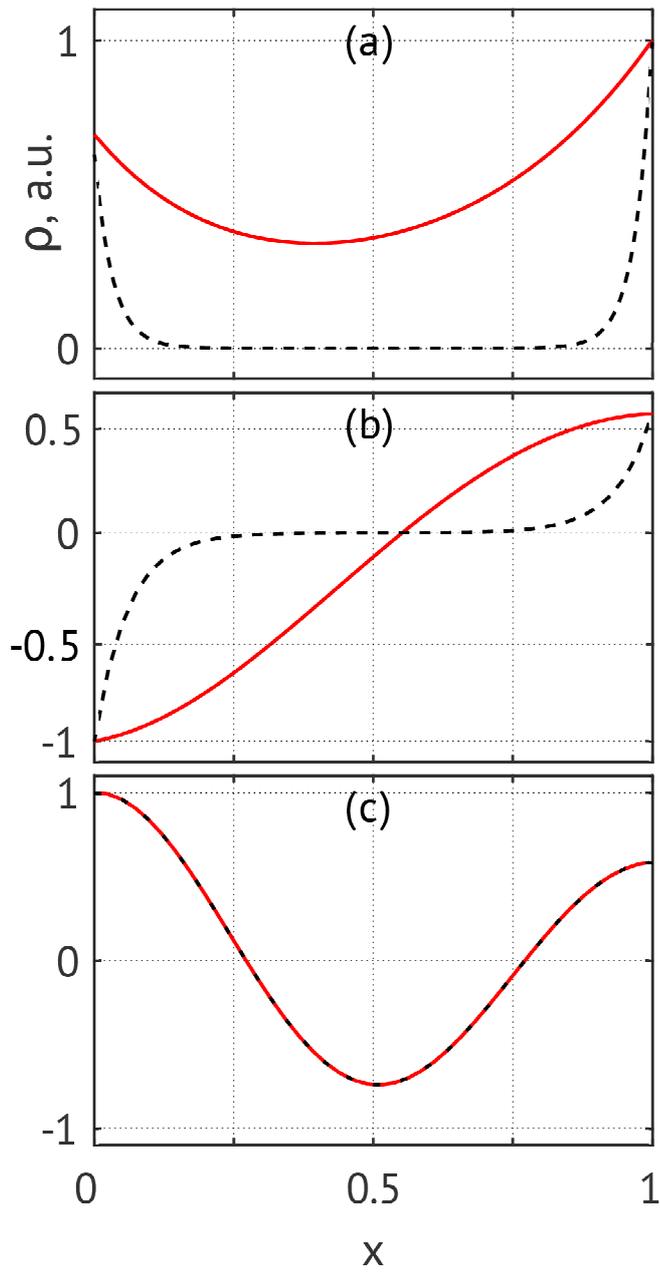

Fig. 3. Radial distributions of the space charge $\rho_n(r)$ at the given polar angle $\vartheta$ for the first three plasmon eigenmodes: (a) $n=1$, (b) $n=2$, (c) $n=3$ in $C_{60}$ ($u=0.314$, solid curves) and in the same shell without polarization nonlocalty ($u=0.031$, dashed curves).

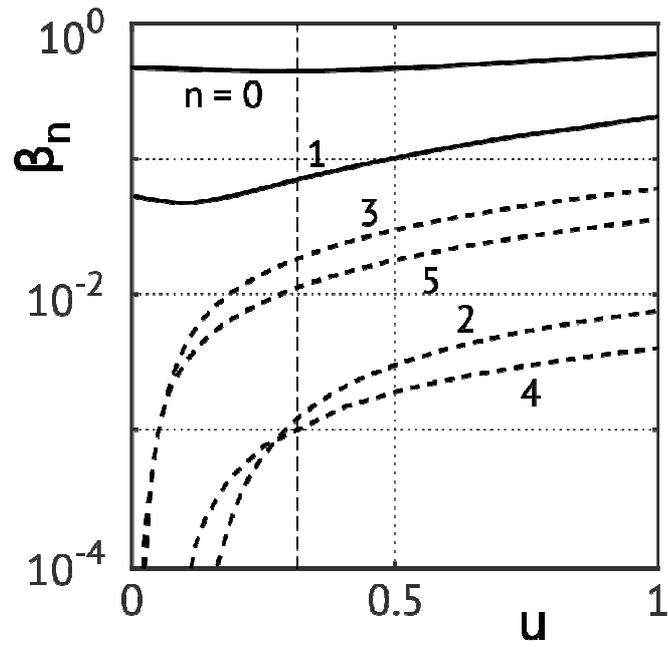

Fig. 4. Excitation factor $\beta_n$ for the first six plasmon modes of $C_{60}$ ($n = 0,1,2,3,4,5$) as a function of the nonlocality parameter $u = V_F/(\omega_p l)$.

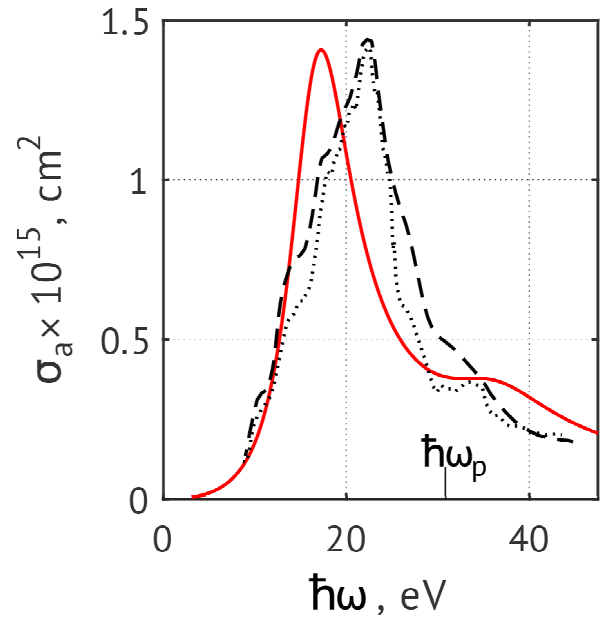

Fig. 5. Frequency dependences of the $C_{60}$ absorption cross-section calculated on the basis of our model (solid curve) and obtained in the experiments described in Ref. 3 (dashed and dotted curves correspond to the experiments fulfilled with $C_{60}$ in the gas and solid phases, respectively).